\documentclass[twocolumn,prl,aps]{revtex4}
\usepackage{graphicx}
\usepackage{epsfig}
\usepackage{amsmath,amssymb,revsymb}
\usepackage{latexsym}

\def\duzomniejsze{<\kern-.7mm<}
\def\duzowieksze{>\kern-.7mm>}

\def\textbf#1{{\bf #1}}
\def\beq{\begin{equation}}
\def\eeq{\end{equation}}
\def\be{\begin{equation}}
\def\ee{\end{equation}}
\def\ben{\begin{eqnarray}}
\def\een{\end{eqnarray}}
\def\beqa{\begin{eqnarray}}
\def\eeqa{\end{eqnarray}}
\def\eea{\end{array}}
\def\bea{\begin{array}}
\newcommand{\bei}{\begin{itemize}}
\newcommand{\eei}{\end{itemize}}
\newcommand{\bee}{\begin{enumerate}}
\newcommand{\eee}{\end{enumerate}}

\def\tr{{\rm Tr}}

\def\ra{\rangle}

\def\>{\rangle}
\def\<{\langle}

\def\ot{\otimes}
\def\rab{\rho_{AB}}

\def\disone{D^\to}
\def\dis{D}
\def\unc{\Upsilon(A:B)}
\def\entc{E_c}
\newcommand{\ket}[1]{| #1 \rangle}
\newcommand{\bra}[1]{\langle #1 |}
\newcommand{\proj}[1]{\ket{#1}\bra{#1}}

\newtheorem{theorem}{Theorem}
\newtheorem{definition}{Definition}

\begin{document}

\title{Uncommon information (the cost of exchanging a quantum state)}
 
\begin{abstract}


If two parties share an unknown quantum state, one can ask how
much quantum communication is needed for party $A$ to send her
share to party $B$.  Recently, it was found that the number of
qubits which should be sent is given by the conditional entropy.
This quantifies the notion of partial information, and it can even
be negative.
Here, we not only demand that $A$ send her state to $B$,
but additionally, $B$ should send his state to $A$.
Paradoxically, we
find that requiring that the parties perform this additional task can
lower the amount of quantum communication required. 
This primitive, which we call {\it quantum state exchange}, can be used to quantify the
notion of {\it uncommon information}, since the two parties only
need to send each other the parts of their state they don't hold in common.  
In the classical case, the concept of uncommon information
follows trivially from the concept of partial information.  We find that
for quantum states, this is not so.  We prove upper and lower bounds
for the uncommon information and find optimal protocols for several
classes of states.
\end{abstract}
\author{Jonathan Oppenheim} 
\affiliation{Department of Applied Mathematics and Theoretical Physics, University of Cambridge U.K.}
\author{Andreas Winter}
\affiliation{Department of Mathematics, University of Bristol, Bristol BS8 1TW, U.K.}
\maketitle

We now understand information in operational terms.  We quantify it in terms of the amount of communication required to
convey messages.  For classical messages represented by a probabilistic source producing
messages $X$ Shannon showed
that a rate of $H(X)$ bits are required to convey the message, where 
$H(X)=-\sum p_x \log_2 p_x$
is the Shannon entropy \cite{Shannon1948}.
Likewise, for a source producing unknown quantum states with density matrix $\rho_A$,
Schumacher \cite{Schumacher1995} showed that $S(A)$ quantum bits (qubits) are necessary and sufficient
to send the states where $S(A)=-\tr \rho_A \log\rho_A$ is the von Neumann entropy and
we drop the explicit dependence on $\rho$.  We thus see that the operational notion of information corresponds
to calculable quantities.

\begin{figure}
\label{fig:cunc}
\centering
\includegraphics[scale=1.0]{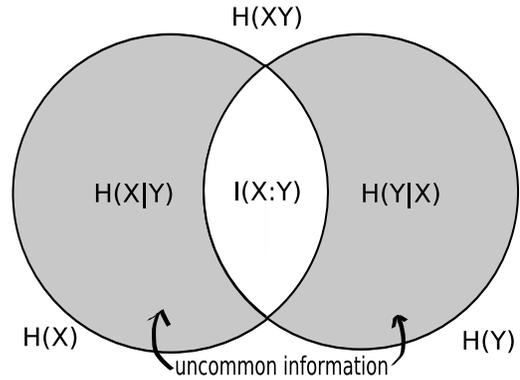}
\caption{A graphical representation of the uncommon information (shaded area)
in classical information theory.
The total information of the
source producing pairs of random variables $X,Y$ is $H(XY)$,
while the information contained in just variable $X$ ($Y$)
is $H(X)$ ($H(Y)$).  The information common to both variables
is the mutual information $I(X:Y)$ (unshaded), while the partial informations
are the conditional entropies $H(X|Y)$ and $H(Y|X)$.  In the quantum case, the 
quantum mutual information $I(A:B)$ can be greater than the total
information $S(AB)$.
To compensate, the partial informations $S(A|B)$ and $S(B|A)$ can be negative. As we show, 
the uncommon information, defined operationally as state-exchange, must be positive in the quantum case.
It thus cannot be the sum of the two partial informations and can not appear on this diagram.}
\end{figure}

Now, if the receiver has some prior information about the messages to be sent,
then generally, less bits (or qubits) need to be sent.  If we represent the receiver's
prior information by the variable $Y$, then the Slepian-Wolf theorem \cite{slepian-wolf}
tells us that $H(X|Y)=H(XY)-H(Y)$ bits will convey the message.  This quantity is
called the conditional entropy, and it gives us a notion of how much 
{\it partial information} needs to be sent if the receiver has some
prior information.  The quantum counter-part of partial information was
recently found by Horodecki and ourselves, \cite{how-merge,how-merge2} through
considering an analogous scenario 
we called {\it quantum state merging}.  Instead
of sharing a random variable $XY$, two parties (named Alice and Bob), share
unknown states from an unknown ensemble with density matrix $\rho_{AB}$.  We then allow free classical communication,
and ask how many qubits
Alice needs to send so that Bob receives her state.  This quantifies
the partial quantum information, and it was shown to be $S(A|B)=S(AB)-S(B)$,
the quantum conditional entropy.  Such a quantity was known previously,
and it had been observed that it can be negative for
entangled states \cite{Wehrl78,HH94-redun,cerfadami}. 
State-merging shows that it has a meaning in terms 
of information.  The fact that it can be
negative then becomes natural -- the conditional entropy quantifies
how many qubits need to be sent from Alice to Bob, and if it is negative,
Alice and Bob gain the potential to send future quantum states
at no cost.  Alice can not only send her state to Bob, but the parties are additionally
left with maximally entangled states which can be used in the future 
to teleport quantum states without using a quantum channel.

Finally, in classical information theory, there is the notion of {\it mutual information} --
the amount of correlation between two variables.  This is given by $I(X:Y)=H(X)+H(Y)-H(XY)$, which 
has the operational meaning as the rate at which messages can be reliably sent through a channel
which takes $X$ to $Y$ (after maximizing $I(X:Y)$ over inputs)~\cite{Shannon1948}. 
The building blocks of classical
information can thus be represented by the Venn diagram of Figure 1.  We will see that
in quantum information theory, the Venn diagram is completely inadequate for representing
the basic building blocks of the theory, and in fact, even entropies appear inadequate.  
Already, the analogous quantity for the mutual information is
slightly less clear.  One analogous task is the sending of 
quantum states through a noisy quantum
channel, which can be done at a rate equal 
to a quantity called the coherent information $I(A\ra B)= S(B)-S(AB)$.
This quantity is asymmetric unlike the quantum mutual information $I(A:B)=S(A)+S(B)-S(AB)$
(which can also be seen as a measure of total correlations -- both classical
and quantum \cite{compl}).
Other quantities which might be seen as giving meaning to the notion of shared (common) correlations in quantum states
including the entanglement cost $\entc$ and the distillable entanglement $\dis$ 
(how many maximally entangled states are required to create a shared state or are obtainable 
from it).  The last one is a natural measure of pure quantum common information as classical
communication is taken to be free for this task. 

Here, we consider a concept which is complementary to the mutual
information -- we thus call it the {\it uncommon information} (since mutual information is sometimes refered to 
as {\it common information}.
In the classical case, we can 
quantify it by considering two parties,
and ask how much classical communication is required for them to exchange their messages.
I.e. if Alice has $X$ and Bob $Y$, how much communication do they need for Bob to get $X$ and
Alice to get $Y$.  This naturally and operationally defines the notion of uncommon information, since they
will have to transfer to each other the parts of their message which the other party
doesn't know about.  It is also a common communication primitive -- most conversations involve
exchanging information.
The solution to this problem follows immediately by application of the 
Slepian-Wolf theorem -- Alice sends $H(X|Y)$ bits to Bob, who then has $X$, and Bob then sends
$H(Y|X)$ bits to Alice so that she has $Y$.  This nicely divides the total information $H(XY)$
into two parts, the mutual information $I(X:Y)$, and the uncommon information 
$U(X:Y)\equiv H(X|Y)+H(Y|X)$.  We then see that $S(XY)=I(X:Y)+U(X:Y)$.

We now want to find the appropriate quantum counterpart to the classical uncommon information.
As we did with state merging, we will consider an operational task which is 
analogous to the classical task -- we call it {\it quantum
state exchange}.  Namely, Alice and Bob share unknown states which are emitted from a source characterized
by density matrix $\rab$ -- they want to swap states,
and we ask how many qubits they need to send in total, while allowing classical communication
for free (since we are interested in isolating the quantum part of the information). 
also consider the case 
Unlike the classical case, where the Slepian-Wolf theorem allows one to quickly
solve message exchange, we will see that one cannot use state merging to
solve this problem -- the situation is completely different.  Indeed this must be
so -- the quantity $S(A|B) + S(B|A)$ can be negative, and if it gave the rate for state exchange,
Alice and Bob would be apply to continually exchange their states, generating an arbitrarily large
amount of pure entanglement.  In essence we will see that, quantum state merging does not
solve the problem of state exchange due to the no-cloning theorem \cite{cloning}.  
 $S(A|B) + S(B|A)$ thus appears to have no physical or operational information-theoretic meaning -- in sharp 
contrast to the classical case.

In the remainder of this article, we will more formally define the notion of quantum state
exchange, and then provide several protocols and solve some examples.  We then provide an
upper bound for the quantum uncommon information $\unc$ 
given by $\entc(R:A)+\entc(R:B)$
where the system $R$ purifies $\rab$, i.e. there is a pure state
$\ket\psi_{ABR}$ and $\tr_R \ket\psi\bra\psi_{ABR}=\rab$.  $S(AB)$ is also
proven to be an upper bound.  $\entc$ is the amount of pure state
entanglement needed to create a state between two parties \cite{BDSW1996,cost}.
We shall then prove
a lower bound on $\unc$ given by $\disone(R\>A)+\disone(R\>B)$ where $\disone(R\ra T)$
is the one way distillable entanglement with classical communication from
$R$ to $T$ only.  Another provable lower bound is
$\max_{\Lambda}\left[S(AV)-S(BV)\right]$ where the maximization is over channels
$\Lambda: R \longrightarrow V$. Stranglely, $S(A|B)+S(B|A)$, the minimum rate for Alice to
send her state to Bob, plus the minimum rate for Bob to send to Alice is not a lower bound,
and we give examples of states for which $S(A|B)$ [or $S(B|A)$] can be very large while the rate
for state exchange is small. 

As with Schumacher compression, we consider the case of sources producing 
unknown states, and we know the statistics of the source only through the
density matrix (extension to the case of unknown sources is
also possible \cite{jozsaHHH98}), but we make no assumptions about the ensemble of
states which may be emitted by the source -- the states are unknown. 
We consider two separated parties, in possession of $n$ copies of state $\rho_{AB}$. A faithful
protocol is one which works with high probability, averaged over all possible unknown states
in an ensemble.  An elegant reformulation of this is to consider a reference system
$R$, and total pure state $\ket\psi_{ABR}$, and define success of the protocols by
demanding that after Alice's state has been transferred to Bob's site, the total
state $\ket\psi_{ABR}$ should be virtually unchanged.  More formally:  

\begin{definition}
A faithful {\bf state merging} protocol  
from Alice to  Bob is an operation that transforms the 
state $\psi_{ABR}^{\ot n}$  
into state $\rho^{merg}_{ABB'B''R}$ such that  for large $n$ 
\be
F(\rho^{merg}_{ABB'B''R},\phi_{AB}\ot\psi^{\otimes n}_{B'B''R})\to 1
\label{eq:def-merg}
\ee
where $\psi_{B'B''R}$ is equal to the original state $\psi_{ABR}$ 
if we substitute $A\to B'$ and $B\to B''$, and the state  $\phi_{AB}$ 
is arbitrary. The fidelity $F(\rho,\sigma)=\tr(\sqrt{\sqrt{\sigma}\rho\sqrt{\sigma}})$, and
subsystems $B$,$B'$,$B''$ are at Bob's site.  Allowing classical communication for
free, the {\bf partial information} is the rate amount of pure entanglement needed to achieve
state merging (taking into account the entanglement left behind in the form
of $\phi_{AB}$).
\end{definition}
Typically, the state $\phi_{AB}$ will be the possible distilled pure entanglement which
might be gained after the protocol.  
Similarly, 
\begin{definition}
A faithful {\bf state exchange} protocol  
between Alice to  Bob is an operation that transforms the 
state $\psi_{ABR}^{\ot n}$  
into state $\rho^{ex}_{ABR}$ such that  for large $n$ 
\be
F(\rho^{ex}_{ABR},\psi_{BAR}^{\otimes n})\to 1
\label{eq:def-ex}
\ee
where $\psi_{BAR}$ is equal to the original state $\psi_{ABR}$ 
if we exchange $A$ with $B$ (i.e. perform the swap operation). 
Allowing classical communication for free, the {\bf uncommon information} $\unc$ is
the minimum rate of pure entanglement required to achieve state exchange.
\end{definition}

State exchange for the completely mixed density matrix is equivalent to
the swap operation and it was shown that two bits of pure entanglement
are necessary and sufficient to perform swap on two qubits \cite{ejpp2000,clp-2000}.
However, we will now see that if some correlations exist between the two systems,
and we have many copies of the state, 
we can actually do better.
We first look at upper bounds for $\unc$, based on specific protocols for
state exchange:

\noindent {\bf Merge-and-send protocol:} Alice merges her state with Bob at a cost of $S(A|B)$, who then
sends his state to Alice at a cost of $S(B)$.  The total rate $R_{ms}$ is thus $S(AB)$.  Here
we see how the no-cloning theorem enters into the situation.  Alice is able to merge her state
with Bob's, taking advantage of the fact that he has some prior information (the state $\rho_B$).
She thus sends her state at the lowest possible rate.
However, once she has sent her state, she is left with nothing (unlike in the classical case
where she can make a clone of her message), and Bob must send at the maximum rate of $S(B)$.
Since individually, each party's partial information can be negative, it is the no-cloning theorem
which prevents the rate of state exchange from being negative.

\noindent{\bf Double-copy protocol:}  Each party coherently copies their state to an ancilla:
$A\rightarrow A'A''$, $B\rightarrow B'B''$ in some known basis which gets
optimized.  I.e. Alice applies the operation 
$\ket{a}_A\ket{0}_{A'}\ket{0}_{A''}\rightarrow \ket{0}_A\ket{a}_{A'}\ket{a}_{A''}$ for
some basis $a$, and similarly Bob performs this operation in some basis $b$ of his choice.
Then:
\begin{enumerate}
\item Alice merges one of her copies of her state to Bob at a cost of $S(A'|B)$
\item Bob merges one of his copies with Alice at a cost of $S(B'|A'')$
\item Alice merges her second copy with Bob at a cost of $S(A''|A'B'')$
\item Bob merges his second copy with Alice, costing $S(B''|B')$ 
\end{enumerate}
The total cost gives the following achievable rate $R_{dc}$ for this protocol
\beqa
R_{dc}&=&\min_{ab}\left(S(A'B)+S(AB')-S(A')-S(B')\right) 
\label{eq:twomentropy}
\\
 &=& \min_{ab}\left(\sum_a p_a S(\sigma_B^a)+\sum_b p_b S(\sigma_A^b)\right) \nonumber\\
 &=&  E_f(A:R)+E_f(B:R)
\label{eq:dcrate}
\eeqa
where the minimization is taken over bases $a$, $b$,
and $E_f$ is the entanglement of formation \cite{BDSW1996}; regularisation
 (optimising over many copies of input state)
leads to the better upper bound by the sum of the entanglement costs \cite{cost}.
The states $\sigma_A^b$  $(\sigma_B^a)$ are
those that would be induced on $A$ $(B)$ after a measurement on $B$ ($A$) with outcomes
$b$ $(a)$ and probabilities $p_b$ ($p_a)$.  The measurement basis is the same
as the basis $\ket{a}$ and $\ket{b}$ chosen in the initial copying step.
The second equality of Equation (\ref{eq:dcrate}), just comes from the fact
that $S(A')$ and $S(B')$ are the same as the classical entropy $H(\{p_a\})$ and $H(\{p_b\})$.
The third equality comes from the fact that a measurement on system $A$ which minimizes
the entropy of system $B$ (conditioned on the outcomes of measurements), can alternatively
be thought of as a measurement which produces pure states $\ket{\psi^a}_{BR}$ -- i.e. it is
a decomposition of $\rho_{BR}$ into pure states  $\ket{\psi^a}_{BR}$ with minimal total 
entanglement.

Expression (\ref{eq:twomentropy}) 
can also
be interpreted as the sum of two 
classical-quantum conditional entropies, each of the form
\beq
SH(A|B)\equiv \inf_\Lambda (S(A \Lambda(B))-S(\Lambda(B))) .
\eeq  
I.e., conditional entropies obtained after applying the decohering map
$\Lambda$ on the conditioning system as is done in the protocol.

Note also that we can express this rate \cite{koashi-winter-2003} in terms of a measurement of classical
correlations, the Henderson-Vedral quantity  \cite{HendersonVedral} which we regularize $C^\infty_{HV}$
\beq
R_{dc}=S(A)+S(B)-C^{\infty}_{HV}(A\>B)-C^{\infty}_{HV}(B\>A) .
\eeq
$C^\infty_{HV}$ is operationally equal to the one-way distillable common randomness 
\cite{DevetakW03-common}.
Given that we have here a very simple protocol involving only four rounds, it seems possible that
a more complicated protocol with many rounds may be related not to $C^\infty_{HV}$
but perhaps the classical information deficit $\Delta_{cl}$ \cite{compl}.

\noindent {\bf Modified double copy protocol:}  One can modify the preceding protocol slightly,
by not required the two parties to perform a complete copying operation.  Each can divide their states
into parts which are copied, and parts which are merged.  We will see that this can be better. 

The double copy protocol is optimal for the so-called classical states,
i.e. states of the form
\beq
\varrho^{cl}_{AB}=\sum p_{ab} \proj{a}\ot \proj{b}
\eeq
since for these states one can copy in the local eigenbasis without changing any of the entropies.
One can thus achieve a rate of $S(A|B)+S(B|A)$ which is optimal due to the lower bound of 
Theorem \ref{thm:lower} proven below. 
The modified double copy protocol appears to be optimal for one-sided classical states, for example
those of the form 
\beq
\varrho^{acl}_{AB}=\sum p_a \proj{a}\ot\sigma^a_{B} 
\eeq
with the classical part on Alice's side.  In this case, Alice copies and merges one
of her copies with Bob who then merges his entire state with Alice's second copy.
Alice then merges this second copy, again achieving the 
rate of $S(A|B)+S(B|A)$.  That $S(B|A)$ qubits need to be sent from Bob follows from Theorem 
\ref{thm:lower} proven below and we suspect that $S(A|B)$ bits also needs to be sent from Alice.

We finally mention another protocol: {\bf do nothing}.  This is possible
(and clearly optimal) for states which
are supported on only the symmetric (or antisymmetric) subspace of Alice and Bob,
or are locally equivalent to such states.
For symmetric or antisymmetric states
on $AB$, $\ket\psi_{ABR}=\pm \ket\psi_{BAR}$, and thus nothing needs to be done
since a global phase does not influence the three-party density operator.
In particular, all pure states on $AB$ have $\unc=0$, as one would expect:
they are fully correlated in the Schmidt
basis, and thus contain no uncommon information.
Surprisingly, neither $S(A|B)$ nor $S(B|A)$ are zero for such
(anti-)symmetric states, thus if the task
is for Alice to send her state to Bob, she needs $S(A|B)$ bits of entanglement,
while if we demand that they perform the additinal task of Bob sending his
state to Alice, the task becomes easier, and no quantum or classical communication
needs to be exchanged.  In the classical case, this of course never happens.


By exhibiting particular protocols, we thus obtain
\begin{theorem} 
\label{thm:upper}
The uncommon information $\unc$ is upper-bounded $\unc\leq S(AB)$
and $\unc\leq E_c(A:R)+E_c(B:R)$.
\end{theorem}

We now turn to lower bounds, and will prove
\begin{theorem}
\label{thm:lower}
The uncommon information $\unc$ satisfies the following lower bounds:
\begin{align*}
  \unc &\geq\disone(R\>A)+\disone(R\>B) \text{ and} \\
  \unc &\geq \max_{\Lambda}\left[S(BV)-S(AV)\right],
\end{align*}
where the maximization
is over channels $\Lambda: R \longrightarrow V$.
\end{theorem}

The proof of Theorem \ref{thm:lower} is straightforward -- for the first inequality,
we imagine $R$ as a referee who
will check to see whether the output state $\rho^{ex}_{ABR}$
is close in fidelity to $\ket\psi_{ABR}$.
Before Alice and Bob begin the protocol, the referee performs
one way distillation with Alice or Bob by
performing local operations on her state.  To distill maximally entangled states 
she would normally communicate with
the other party, but this isn't necessary -- it is only used to tell Alice or Bob which parts of
their state contain the distilled entanglement.  From the referee's perspective she holds
$\disone(R\>A)$ (or $\disone(R\>B)$) bits of pure entanglement 
with $A$ (or $B$).  
Imagine she distills entanglement (i.e. state $\psi^+$) with Alice.  
Then clearly $\psi^+$ on $RA$ must be transferred by Alice to Bob, since the
referee can check after completion of the protocol by asking Bob for the appropriate
bits, and checking the fidelity of the subsystem of $\rho^{ex}_{ABR}$ which contains  $\psi^+$.
However,
%
%
Alice and Bob perform their protocol before they know which
party the referee distilled with, and thus from their point of view,
pure state entanglement with $R$ may exist on both their
states which needs to be transferred to the other party.  Thus just as Alice needs
to transfer $\disone(R\>A)$ to Bob, Bob also needs to
transfer $\disone(R\>B)$ qubits to Alice in case the referee distilled entanglement with him.

The second inequality comes by imagining dividing $R$ into two parts, $E$ and $V$
(which is equivalent to a channel with $E$ treated as the channel's environment),
and giving $\rho_V$ to Alice and $\rho_E$ to Bob.  Before the protocol, the entanglement is 
$S(AV)$ and after the state exchange the entanglement is $S(BV)$.  Since entanglement cannot
increase more than the number of qubits exchanged, the difference between final
and initial entanglement is a lower bound on the number of sent qubits.  
Optimizing over splittings of $\rho_R$
gives the required bound.    \hfill $\Box$


Let us turn from uncommon information to the notion of common information (which can be taken as
a more general notion than mutual information \cite{othercommon}).  State exchange
considerations suggest that it be given by 
${\cal C}(A:B)\equiv S(AB)-\unc$, i.e. the uncommon information
subtracted from the total information.
Such a quantity is always positive by Theorem \ref{thm:upper},
but is very different from the mutual information (for example, it is zero for pure states).
Part of the reason for this is that mutual information measures the correlations between
the two parties, while the common information quantifies how much information {\em about 
a reference system} the two
parties share in common.  It is thus zero for pure states because a pure state has no
information about the reference system, while it is maximal for the 
symmetric states, where all the information is common.
It would be interesting to explore this notion of common
information further, especially compared with other notions 
such as the mutual information, coherent information, distillable
entanglement and entanglement of formation.

\begin{acknowledgments}
We thank Charlie Bennett for suggesting the nickname for the uncommon information.
JO acknowledges the support of 
the Royal Society, the Newton Trust, and EU grant
PROSECCO (IST-2001-39227). AW was supported by EU grant RESQ
(IST-2001-37559), the U.K. Engineering and Physical 
Sciences Research Council's ``QIP IRC'', and a University of
Bristol Resesarch Fellowship.
\end{acknowledgments}

\bibliographystyle{apsrev}


\end{document}